\documentclass[12pt]{article}
\usepackage{epsfig}

\begin{document}

\title{Cooperation, punishment, emergence of government and the tragedy of
authorities}
\author{R. Vilela Mendes \thanks{%
also at IPFN, Instituto Superior T\'{e}cnico, Av. Rovisco Pais 1, 1049-001
Lisboa} \thanks{%
e-mail: vilela@cii.fc.ul.pt} \\
CMAF, Complexo Interdisciplinar,\\
Av. Gama Pinto 2, 1649-003 Lisboa, Portugal \and Carlos Aguirre \thanks{%
e-mail: carlos.aguirre@uam.es} \\
Escuela Polit\'{e}cnica Superior, \\
Universidad Autonoma de Madrid, Campus de Cantoblanco, \\
Ctra de Colmenar Km 16, 28049 Madrid, Spain}
\date{}
\maketitle

\begin{abstract}
Under the conditions prevalent in the late Pleistocene (small
hunter-gatherer groups and frequent inter-group conflicts), coevolution of
gene-related behavior and culturally transmitted group-level institutions
provides a plausible explanation for the parochial altruistic and
reciprocator traits of most modern humans. When, with the agricultural
revolution, societies became larger and more complex, the collective nature
of the monitoring and punishment of norm violators was no longer effective.
This led to the emergence of new institutions of governance and social
hierarchies. Likely, the smooth acceptance of the new institutions was
possible only because, in the majority of the population, the reciprocator
trait had become an internalized norm. However the new ruling class has its
own dynamics which in turn may lead to new social crisis. Using a simple
model, inspired on previous work by Bowles and Gintis, these effects are
studied here.
\end{abstract}


\section{Introduction}

It is a fact that humans are a highly cooperative species. Cooperative in
helping each other, cooperative in achieving material and intellectual
achievements unmatched by other species, but also cooperative in war and
genocide. From the biological point of view, human cooperation is an
evolutionary puzzle. Unlike other creatures, humans cooperate with
genetically unrelated individuals, with people they will never meet again,
when reputation gains are small or absent and even engage in altruistic
punishment of defectors. These patterns of cooperation cannot be explained
by kin selection, signalling theory or reciprocal altruism. The idea that
group selection might explain this behavior goes back to Darwin himself who,
in chapter 5 of the \textit{``Descent of man and selection in relation to
sex''}, states that ``\textit{... an increase in the number of well-endowed
men and an advancement in the standard of morality will certainly give an
immense advantage of one tribe over another.}'' However, this idea felt in
disrepute because evolution does not pitch groups again groups , nor
individuals against individuals, but genes against genes. Then, a ``selfish
gene'' analysis makes the altruistic good-of-the-group outcome virtually
impossible to achieve. In particular because the late Pleistocene groups of
modern man were not believed to be genetically sufficiently different to
favor group selection. Therefore, human cooperation remained an evolutionary
puzzle.

In recent years Bowles, Gintis and collaborators \cite{Bowles2} \cite%
{Bowles2a} \cite{Bowles3} \cite{Bowles4} revived the group selection idea by
showing that the particular environment and type of the hunter-gatherer
groups of the late Pleistocene (which corresponds to about 95\% of the
evolutionary time of modern man) were such that a multilevel evolutionary
dynamics involving gene-culture coevolution could account for the
development of the cooperative altruistic trait which they call \textit{%
strong reciprocity}. The cost of group beneficial behavior to an individual
would be limited by the emergence of group-level social norms. On the other
hand, even in the absence of these group-level norms, group selection
pressures would support the evolution of the cooperative-altruistic
punishment trait if intergroup conflicts were very frequent. Egalitarian
practices among ancestral humans reduced the force of individual selection
against altruists, while frequent warfare made altruistic cooperation among
group members essential to survival. That is, parochial altruism and warfare
could have coevolved. Furthermore they developed simple mathematical models
that gave quantitative support to their ideas.

I think that the analysis of Bowles and Gintis provides a convincing picture
of the genesis of the cooperative nature of humans and their culture. The
human capacity for social norm building and for the cultural transmission of
learned behavior allowed altruistic other-regarding preferences to
proliferate. But it also suggests that the other-regarding preferences that
we inherited from primeval man are partly cultural not purely genetic.
Therefore liable to change at a much faster pace than if they were purely
genetic. A natural question is what is happening to this human trait (that
presumably developed during a period of 190000 years) in the short time
(10000 years) since the end of the Pleistocene. Using a simple version of
the Bowles-Gintis model I have analyzed in \cite{Vilela1} the evolution of
the reciprocator trait in a situation where the size of the society and the
degree of clustering precludes the collective nature of rule violator
monitoring. Both an agent-based and a mean-field model were used. The main
conclusion was that in this situation the reciprocator trait would not be
evolutionary stable.

Historically one knows that such transition from the small hunter-gatherer
groups to larger sedentary population groups occurred at the time of the
agricultural revolution and that the solution was \textit{the emergence of
government}. That is, a new type of agent (the ruler, the authority) came
into play and replaced the type of egalitarian decision-making that might
have existed before. It is worth noticing that the apparent ease with which
humans accepted this transition of power may have much to do with the
internalization of the reciprocator trait, that above complete freedom
valued the enforcing of social norms.

In the agricultural societies, specialization arose as well as new security
needs and more intense population pressure on limited resources. This tended
to produce greater organization within the community, which in turn led to
social hierarchies, to certain forms of chieftainship and to a whole class
of people with managing roles. The government rulers that the first
agricultural societies accepted were in general priestly figures (the
summerian Ensi) which might have depicted themselves as servants of the
gods, acting in behalf of the community. Here the emergence of organized
religion appears as a norm-enforcing tool, because it is easier for the
ruler to invoke the will of the gods than their own personal preferences. As
time went on, especially because of the creation of other rival city states,
the cities came to rely more and more on military leaders and the ruling
priests gave place to military leaders (Lugal -- the King).

In this paper, using a setting similar to the one in \cite{Vilela1}, I will
study the effect of introducing in the model a new agent representing the
role of the authorities. The collective monitoring and punishment of the
reciprocators will be a decreasing function of the population size in the
social group which is allowed to grow with the average fitness. The need to
introduce authority agents to avoid a \textquotedblleft tragedy of the
commons\textquotedblright , that is, a fitness crisis arising from the
proliferation of the self-regarding agents, is an expected effect. This is
what was called above the \textit{emergence of government}. The interesting
question is that the dynamics of the authority agents may, by itself, lead
to a new fitness crisis which will called a \textit{tragedy of authorities}.
This crisis may or may not be related to the \'{e}lite overproduction crisis
that some authors \cite{Turchin1} \cite{Turchin2} \cite{Korotayev} have
identified. This will be discussed in the final section of the paper.

\section{Emergence of government and the ``tragedy of authorities''}

The basic setting is similar to the one used before \cite{Bowles2a} \cite%
{Vilela1} as far as the type of \textit{public goods activity} is concerned
in a group of $N$ agents, $N$ being in general a function of time. Here
however three types of agents are considered. The first type (R-agents) are
cooperators that also have a monitoring effect on the cooperation of other
agents. The second are self-regarding agents (S-agents) and the third are
purely monitoring agents (A-agents). The labels that were chosen refer to
the names reciprocators (R), self-regarding or shirkers (S) and authorities
(A). The percentages of each one of the types in the population are denoted
by $f_{R}$, $f_{S}$ and $f_{A}$.

Each R or S-agent can produce a maximum amount of goods $q$ at cost $b$
(with goods and costs in fitness units). An S-agent benefits from shirking
public goods work by decreasing the cost of effort $b\left( \sigma \right) $%
, $\sigma $ being the fraction of time the agent shirks. As before, the
following conditions hold 
\begin{equation}
b\left( 0\right) =b,\qquad b\left( 1\right) =0,\qquad b^{^{\prime }}\left(
\sigma \right) <0,\qquad b^{^{\prime \prime }}\left( \sigma \right) >0
\label{2.1}
\end{equation}%
Furthermore $q\left( 1-\sigma \right) >b\left( \sigma \right) $ so that, at
every level of effort, working helps the group more than it hurts the worker.

For $b\left( \sigma \right) $ one chooses \cite{Bowles2} 
\begin{equation}
b\left( \sigma \right) =\frac{2}{2\sigma -1+\sqrt{1+4/b}}-\frac{2}{1+\sqrt{%
1+4/b}}  \label{2.2}
\end{equation}%
which satisfies the constraints (\ref{2.1}).

R-agents never shirk and punish each free-rider at cost $c\sigma $ and
probability $p\left( N\right) $, the cost being shared by all R-agents. For
an S-agent the estimated cost of being punished is $s\sigma $, punishment
being ostracism or some other fitness decreasing measure. Punishment and
cost of punishment are proportional to the shirking time $\sigma $. $c$ is
the reciprocator unit of punishment cost. $s$ is the weight given by an
S-agent to the possibility of being punished. It may or may not be the same
as the actual fitness costs of punishment ($\gamma ,\gamma _{A}$). Each
S-agent chooses $\sigma $ (the shirking time fraction) to minimize the
function 
\begin{equation}
B\left( \sigma \right) =b\left( \sigma \right) +s\left( f_{R}+f_{A}\right)
\sigma -q\left( 1-\sigma \right) \frac{1}{N}  \label{2.3}
\end{equation}%
From the point of view of an S-agent $\left( f_{R}+f_{A}\right) \sigma $ is
the probability of being monitored and punished. The last term is the
agent's share of his own production. The value $\sigma _{S}$ that minimizes $%
B\left( \sigma \right) $ is 
\begin{equation}
\sigma _{S}=\max \left( \min \left( \frac{1}{2}-\sqrt{\frac{1}{4}+\frac{1}{b}%
}+\frac{1}{\sqrt{s\left( f_{R}+f_{A}\right) +\frac{q}{N}}},1\right) ,0\right)
\label{2.4}
\end{equation}

The contribution of each species to the population in the next time period
is proportional to its fitness $\pi _{R}$, $\pi _{S}$ or $\pi _{A}$ computed
from 
\begin{equation}
\begin{array}{lll}
\pi _{R}^{^{\prime }} & = & q\left( 1-f_{A}-f_{S}\sigma _{S}\right)
x-b-cp\left( N\right) f_{S}\frac{N\sigma _{S}}{Nf_{R}} \\ 
\pi _{S}^{^{\prime }} & = & q\left( 1-f_{A}-f_{S}\sigma _{S}\right)
x-b\left( \sigma _{S}\right) -\left( \gamma p\left( N\right) f_{R}+\gamma
_{A}f_{A}\right) \sigma _{S} \\ 
\pi _{A}^{^{\prime }} & = & q\left( 1-f_{A}-f_{S}\sigma _{S}\right)
wx-c_{A}f_{S}\frac{N\sigma _{S}}{Nf_{A}}%
\end{array}
\label{2.5}
\end{equation}%
and $\pi _{R,S,A}=\max \left( \pi _{R,S,A}^{^{\prime }},0\right) $ because
the baseline fitness is zero.

The first term in both $\pi _{R}^{^{\prime }}$, $\pi _{S}^{^{\prime }}$ and $%
\pi _{A}^{^{\prime }}$ is the benefit arising from the produced public
goods. The factors $x$ and $wx$ with 
\[
x=\frac{1}{wf_{A}+1-f_{A}} 
\]
account for the fact that this benefit is the same for R and S-agents but
might be different for A-agents. The second term in $\pi _{R}^{^{\prime }}$
and $\pi _{S}^{^{\prime }}$ is the work effort. The third term in $\pi
_{R}^{^{\prime }}$ and the second term in $\pi _{A}^{^{\prime }}$ represent
the fitness cost of punishment for R and A-agents and the third term in $\pi
_{S}^{^{\prime }}$ the cost incurred by S-agents when they are punished.

The $\gamma $ and $\gamma _{A}$ coefficients code for the severity of the
coercive measures affecting the fitness of S-agents. The last term in $\pi
_{R}^{^{\prime }}$ and $\pi _{A}^{^{\prime }}$ emphasizes the heavy
punishing burden put on R or A-agents when in small number. The factor $%
p\left( N\right) $, a decreasing function of $N$, accounts for the fact that
(as studied at length in \cite{Vilela1}), when a social group grows in size,
the collective nature of monitoring of free-riders becomes increasingly
difficult. Essentially, the punishment probability by R-agents should be a
growing function of the clustering coefficient of the group. Here, for
illustration purposes, one chooses a simple function of $N$%
\[
p\left( N\right) =\sqrt{\frac{1+\delta }{1+\delta \frac{N}{N_{0}}}} 
\]%
$N_{0}$ being some small initial population.

Finally, for the evolution of the population at successive generations, one
chooses a replicator map\footnote{%
A different, incremental, dynamics is sometimes used for the fitness-based
evolution of populations. The replicator map used here provides faster
evolution but qualitatively similar results, up to a renormalization of the
time scale.} 
\begin{equation}
f_{\alpha ,\mathnormal{new}}=f_{\alpha }\frac{\Pi _{\alpha }\left( f\right) 
}{f_{R}\Pi _{S}+f_{s}\Pi _{S}+f_{A}\Pi _{A}}  \label{2.6}
\end{equation}%
$\alpha =R,S,A$.

First one studies the dynamics of R and S-agents alone, keeping $f_{A}=0$.
In this case, using (\ref{2.4}) and (\ref{2.5}), the evolution of $%
f_{R}=1-f_{s}$ corresponds to a one-dimensional map which is illustrated in
Fig.1 for two values of $p\left( N\right) $ ($1,0$ and $0.5$). For $p\left(
N\right) =1$ the map has an unstable fixed point at $A$ ($f_{R}\left(
A\right) \simeq 0.57$), a left-stable fixed point at $B$ ($f_{R}\left(
B\right) \simeq 0.85$) and a continuum of neutral fixed points after that.
For $p\left( N\right) =0.5$ only the neutral fixed points remain. The
neutral fixed points correspond to the situation where S-agents do not shirk
for fear of being punished. For initial conditions smaller than $f_{R}\left(
A\right) $ in the first case or $f_{R}\left( B\right) $ in the second the
population of R-agents is always invaded by S-agents. However the neutrality
of the fixed points means that the population of S-agents is not completely
invaded by the R-agents.

\begin{figure}[htb]
\begin{center}
\psfig{figure=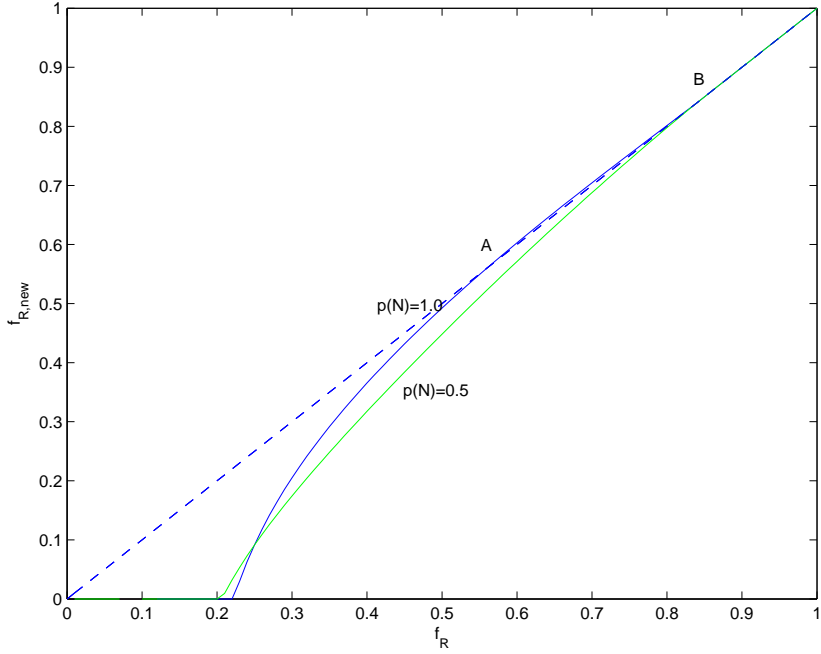,width=11truecm}
\end{center}
\caption{One-dimensional map for the evolution of R-agents corresponding 
to $f_{A}=0,q=2,b=1,c=0.1,\gamma=4,s=3,N=20$}
\end{figure}

Next, still keeping $f_{A}=0$, the evolution of the population of R and
S-agents is studied when the population increases in time according to a
global fitness dependent law, chosen as 
\[
N\left( t+1\right) =N\left( t\right) e^{\beta \pi } 
\]
with $\pi =\sum_{\alpha }f_{\alpha }\pi _{\alpha }$.

In Fig.2 are displayed the results for a time-evolution starting from $%
N_{0}=20$, $f_{R}=0.7$, $f_{S}=0.3$. In the upper left plot the percentages $%
f_{R},f_{S}$ and $f_{A}$ ($f_{A}=0$ in this case) of each agent type are
displayed as the distances to the three sides of a triangle. One sees that
as long as the population ($N$) remains small the monitoring effects of
R-agents controls shirking ($\sigma $) by the S-agents and, as a result,
their percentage ($f_{R}$) and fitness ($\Pi _{R}$) increases as well as the
average fitness of the group. However, with further population growth the
punishment probability ($p\left( N\right) $) of shirkers decreases leading
for a while to a higher degree of shirking ($\sigma $) and higher fitness ($%
\Pi _{S}$) and percentage ($f_{S}$) of S-agents. But because S-agents with
high $\sigma $ produce much less goods, finally the fitness of all agents
decreases and the group collapses. This is the well-known \textit{tragedy of
the commons}, here induced by the fact that monitoring of the public goods
behavior of the agents cannot be a fully collective activity in a large
society.

\begin{figure}[htb]
\begin{center}
\psfig{figure=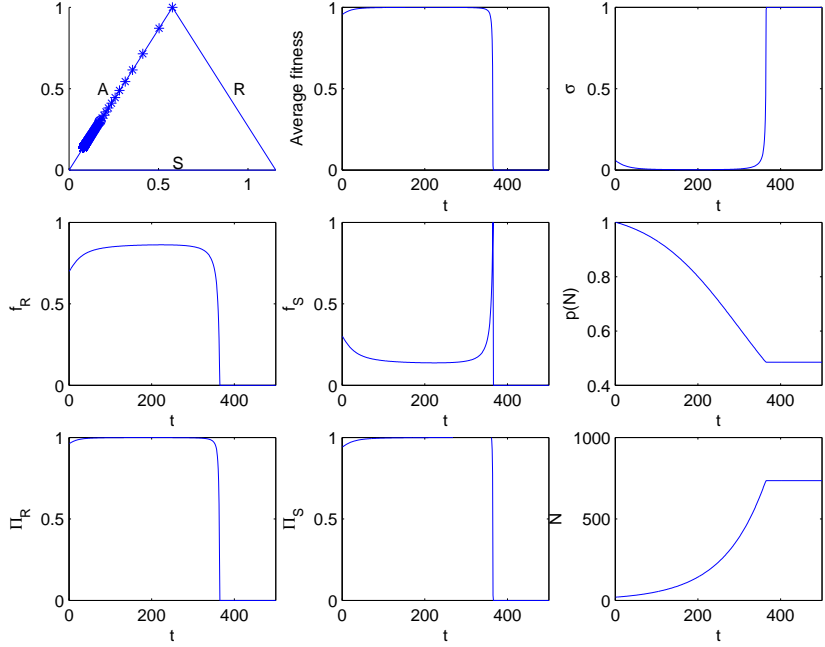,width=11truecm}
\end{center}
\caption{Time evolution of R- and S-agents with $f_{A}=0,q=2,b=1,c=0.1,
\gamma =4,s=3,N_{0}=20$}
\end{figure}

It is then natural that a population group whose success is based on
cooperation and control of selfish behavior, would recognize the need,
beyond a certain population level, to assign the control and punishing role
to specialized agents, with extra power and authority. This is what one
might call the \textit{emergence of government}. In the model, one now
starts from the same initial conditions, but when $f_{R}$ reaches a value
below $0.5$ unfreeze the dynamics of A-agents, imposing however, for the
moment, the constraint that $f_{A}$ should not exceed $0.2$ and, to isolate
the effect of the A-agents, the population is assumed to be constant after
that moment. The result is shown in Fig.3.

\begin{figure}[htb]
\begin{center}
\psfig{figure=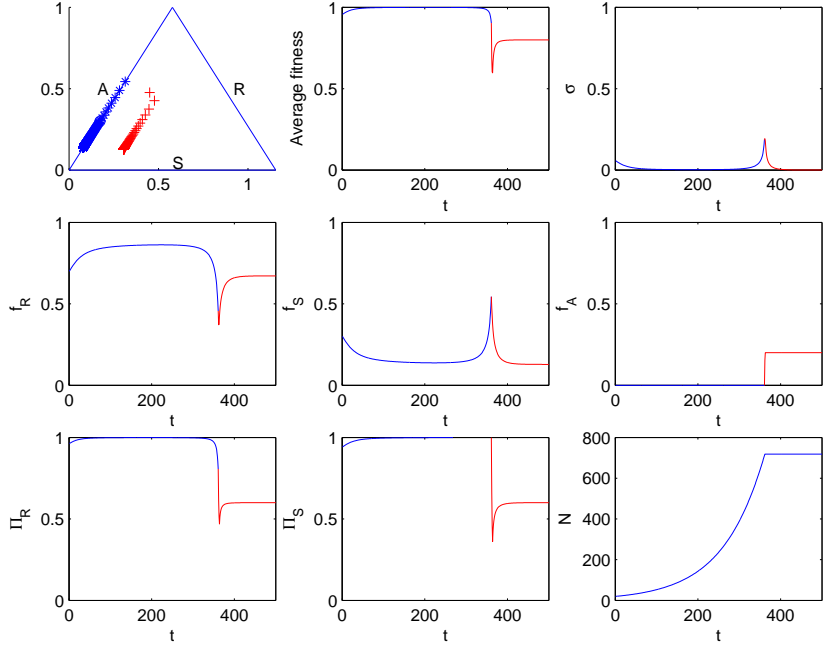,width=11truecm}
\end{center}
\caption{Time evolution with the three
types of agents but $f_{A}\leq 0.2$. ($q=2,b=1,c=0.1,s=3,c_{A}=0.45,
\gamma=4,\gamma_{A}=11$)}
\end{figure}

The outcome is rather satisfactory. After the unfreezing of the $f_{A}$
dynamics the percentage of R-agents still decreases for a while, but then it
stars to grow and the group stabilizes at an high level of average fitness.

Notice that the growth of the number of A-agents is rather fast. The reason
is that as soon as they start controlling the behavior of the S-agents, both 
$\sigma $ and $f_{S}$ decrease, therefore greatly increasing the fitness of
the A-agents, because they benefit from the goods produced without incurring
the cost of control because there is almost nothing to control anymore. If
one now removes the $0.2$ bound on $f_{A}$ (Fig.4) the A-agents population
continues to grow but, because they produce no goods, the average fitness
finally decreases to zero as the group collapses. This is a crisis of a
different type that one might call the \textit{tragedy of authorities}. What
this means for actual societies will be discussed later.

A very similar effect is obtained if, while keeping $f_{A}$ bounded, one
allows $w$ to grow with the fitness of A-agents. That is, allowing the share
of goods allotted to A-agents to grow.

\begin{figure}[htb]
\begin{center}
\psfig{figure=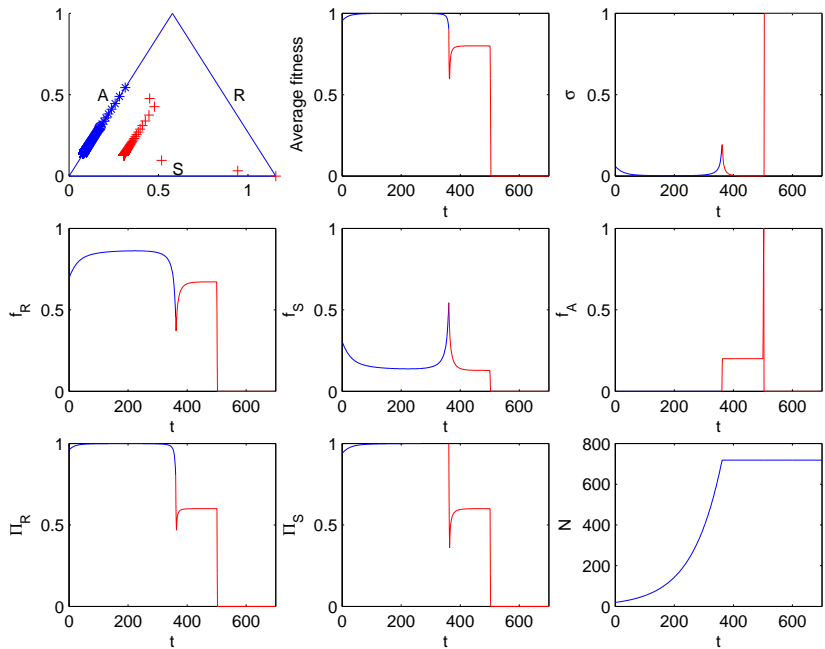,width=11truecm}
\end{center}
\caption{Time evolution with the three
types of agents and $f_{A}$ allowed to grow above $0.2$ after time $500$.
($q=2,b=1,c=0.1,s=3,c_{A}=0.45,\gamma =4,\gamma _{A}=11$}
\end{figure}

Here I would only like to emphasize the delicate nature of the balance
between the several agents in a viable society and the emergence of what
seem to be universal features in the human social evolution. Cooperation is
at the root of success in human groups. However a natural, perhaps
biological, tendency of humans to minimize effort and to maximize benefits
requires that a certain amount of control of shirking is required. This led
some humans to internalize the idea that shirkers should be controlled.
Apparently, it is the societies where more humans adopted this norm that
were the most successful. When, after the agricultural revolution the human
groups became larger, collective control became more difficult. Then, the
evolved acceptance of social norms led naturally to the acceptance of
government as a specialized body. However the dynamics of the authority
agents may, by itself, lead to a new fitness crisis.

\section{Remarks and conclusions}

Stylized mathematical models, both in natural and human sciences, are not
intended to take care of all the details that each particular system
possesses. Rather, they are intended to extract general features or
universal mechanisms, if any, that rule the dynamics of the system. Then, of
course, the detailed characteristics that each physical system or society
has, will determine the time scales and intensity of the universal features.

The general features that may be extracted from this and previous works are:

1) Under the conditions prevalent in the late Pleistocene, that is, small
population groups, frequent inter-group conflicts and a species with the
capacity for norm enforcing and cultural transmission of learned behavior,
the reciprocator trait may become dominant although, in general, not
completely invasive of the self-regarding type.

2) In a large population, monitoring of public goods behavior cannot be a
fully collective activity, rather being the chore of those in close contact
with the free-riders. Because punishment of free-riders requires a local
consensus among reciprocators, the clustering nature of the society would
play an important role in the maintenance and evolution of the reciprocator
trait. Although large human societies tend to be \textquotedblleft small
worlds\textquotedblright\ in the sense of short path lengths they do not
necessarily maintain a high degree of clustering. Therefore norm monitoring
and enforcing requires new special institutions of governance. However, the
new institutions bring with them social hierarchies, which imply
inequalities. Therefore, acceptance of the new institutions is only possible
if in the majority of the population the reciprocator trait had become an
internalized norm.

3) The evolutionary dynamics of the agents associated to governance, that is
the ruling class, may by its proliferation or by assigning to itself a
higher share of the production (an high $w$ factor in the model of Section
2) provoke a decrease of the average fitness, a crisis or even a collapse of
the society. This is what has been called here the \textit{tragedy of
authorities}. Some authors \cite{Turchin1} \cite{Turchin2} \cite{Korotayev}
have studied the historical effects of \textquotedblleft \'{e}lite
overproduction\textquotedblright\ as generating crisis and revolutions.
However not all cases of \'{e}lite overproduction that they characterize can
be identified with the phenomena of the tragedy of authorities. If \'{e}lite
overproduction is, for example, the proliferation of an aristocratic class,
that under the protection of the ruler lives from the society production
without contributing to it, then it has all the marks of a tragedy of
authorities. But if, instead, \'{e}lite overproduction is associated to an
higher access of the youth to higher education, this is not a tragedy of
authorities. The eventual crisis that may occur in this case results from
the fact that the new educated agents are not incorporated neither in the
productive sector nor as beneficiaries of the society production. Hence it
is not a tragedy of authorities. In fact they are only reacting against an
authority structure that wants to preserve their privileges. Therefore to
associate this two distinct situations under the same \'{e}lite
overproduction label may be quite misleading.

As has been shown in Section 2, the existence of authority agents is
beneficial to society as long as their number and their share of the goods
remains limited. The problem therefore is the old question of \textit{who
controls the controllers}. Democracy is in principle a way to implement
limitations and accountability of the rulers. But even then, nothing is
guaranteed. Economic power easily escapes constraints of democratic control.
And even more subtle effects may occur. For example, through exploration of
the co-evolved parochial feelings of the population, it is easy to erect as
a goal the proliferation of local or regional government structures,
coordinating committees, etc. Layers and layers of control when there is
nothing else to control.

4) Even subtler effects of emergent tragedies of authorities are found
everywhere. The solidary form of collective government of the
hunter-gatherer groups was probably the most successful invention of modern
man, leading to his dominance over other species and even over other
hominids. It was also the most extensively tested of all, lasting for 95\%
of the evolutionary history of modern man. Centralized, professional forms
of government, by comparison, are a very recent development, not always very
successful. Hence, it could be rationally expected that, whenever
applicable, "community government" would be used. In fact and except in very
rare cases this is not so. Instead, centralized forms of government tend to
migrate to all local levels carrying with them the kind of political
party-oriented issues, which are not necessarily the most relevant at the
local community level.

5) Evolutionary stability of the reciprocator trait is very much dependent
on social norms and transmission of culture. Therefore it is a trait that
depends as much on genetics as on culture. Some direct evidence of this
comes from the fact that experimental games played by adults and young
children have different results. Culturally-inherited traits may have a much
faster dynamics than gene-based ones. Therefore if the reciprocator trait
has a high cultural component, it is critical to understand how modern
society might be acting on or modifying it. A considerable loss of
cooperative behavior might change society in many unexpected ways. Could
less altruism come along with less hostility to strangers? If contemporary
man is becoming more Homo Economicus maybe it would not be necessary to
rewrite the classical economy books.

\end{document}